\documentclass[12pt,preprint]{aastex}

\newcommand{\mincir}{\ \raise -2.truept\hbox{\rlap{\hbox{$\sim$}}\raise5.truept
        \hbox{$<$}\ }}
\newcommand{\magcir}{\ \raise -2.truept\hbox{\rlap{\hbox{$\sim$}}\raise5.truept
 	\hbox{$>$}\ }}

\shorttitle{High-redshift QSOs in the GOODS Survey}
\shortauthors{Cristiani et al.}

\begin{document}

\title{The space density of high-redshift QSOs in the GOODS Survey 
\footnote{Based on observations taken with the NASA/ESA Hubble Space
Telescope, which is operated by the Association of Universities for
Research in Astronomy, Inc.\ (AURA) under NASA contract NAS5--26555.}
}

\author{
S.Cristiani\altaffilmark{1,8},
D.M.Alexander\altaffilmark{2,10},
F.Bauer\altaffilmark{2},
W.N.Brandt\altaffilmark{2},
E.T.Chatzichristou\altaffilmark{3},
F.Fontanot\altaffilmark{4},
A.Grazian\altaffilmark{5},
A.Koekemoer\altaffilmark{6},
R.A.Lucas\altaffilmark{6},
P.Monaco\altaffilmark{4},
M.Nonino\altaffilmark{1},
P.Padovani\altaffilmark{6,9},
D.Stern\altaffilmark{7},
P.Tozzi\altaffilmark{1},
E.Treister\altaffilmark{3},
C.M.Urry\altaffilmark{3},
E.Vanzella\altaffilmark{8}
}

\altaffiltext{1}{INAF-Osservatorio Astronomico, Via Tiepolo 11, I-34131 Trieste, Italy, e-mail: cristiani@ts.astro.it}
\altaffiltext{2}{Departmrnt of Astronomy and Astrophysics, Pennsylvania State
University, 525 Davey Lab, University Park, PA 16802}
\altaffiltext{3}{Department of Astronomy, Yale University, PO Box
208101, New Haven, CT06520}
\altaffiltext{4}{Dipartimento di Astronomia dell'Universit\`a, Via Tiepolo
11, I-34131 Trieste, Italy}
\altaffiltext{5}{INAF-Osservatorio Astronomico di Roma, via Frascati 33, I-00040
Monteporzio, Italy}
\altaffiltext{6}{Space Telescope Science Institute, 3700 San Martin
Dr., Baltimore, MD 21218}
\altaffiltext{7}{Jet Propulsion Laboratory, California Institute of Technology, 
Mail Stop 169-506, Pasadena, CA 91109}
\altaffiltext{8}{European Southern Observatory,
K.Schwarzschild-Stra\ss e 2, D-85748 Garching, Germany}
\altaffiltext{9}{ESA Space Telescope Division}
\altaffiltext{10}{Institute of Astronomy, Madingley Road, Cambridge, CB3 0HA, UK}
\begin{abstract}
We present a sample of 17 high-redshift ($3.5 \mincir z \mincir 5.2$)
QSO candidates in the $320$ arcmin$^2$ area of the Great
Observatories Origins Deep Survey,
selected in the magnitude range $22.45 < z_{850} < 25.25$ using deep
imaging with the Advanced Camera for Surveys onboard the {\em Hubble
Space Telescope} and the Advanced CCD Imaging Spectrometer onboard the
{\em Chandra X-ray Observatory}.
On the basis of seven spectroscopic and ten photometric redshifts we
estimate that the final sample will contain between two and four
QSOs with $4 < z < 5.2$.  A dearth of high-redshift,
moderate-luminosity ($M_{145} \simeq -23$) QSOs is observed with
respect to predictions based on
{\em a)} the extrapolation of the $z \sim 2.7$ luminosity function
(LF), according to a pure luminosity evolution calibrated by the
results of the Sloan Digital Sky Survey;
and {\em b)} a constant universal efficiency in
the formation of super-massive black holes (SMBHs) in dark-matter
halos.  Evidence is gathered in favor of a density evolution of the
LF at high redshift and of a suppression of
the formation or feeding of SMBHs in low-mass halos.
\end{abstract}
\section{Introduction}
QSOs are intrinsically luminous and therefore can be seen rather
easily at large distances; however, they are rare, and so finding them
requires surveys over large areas. 
As a consequence, at present, the number density of QSOs at 
high redshift is not well known.
Recently, the Sloan Digital Sky Survey (SDSS) has produced a
breakthrough, discovering QSOs up to $z=6.43$
\citep{Fan03} and building a sample of six QSOs with $z>5.7$.  The
SDSS, however, is sensitive only to very luminous QSOs
($M_{145} \mincir -26.5$) and provides no information about the faint
end of the high-$z$ QSO LF, which is
particularly important to understand the interplay between the
formation of galaxies and super-massive black holes (SMBH) 
and to measure the QSO contribution
to the UV ionizing background \citep{Madau99}.
New deep multi-wavelength surveys like the Great Observatories Origins
Deep Survey (GOODS) provide significant constraints
on the space density of less luminous QSOs at high
redshift. 
Here we present high-$z$
QSO candidates, identified in the two GOODS fields on the basis of
deep imaging in the optical (with {\em HST}) and X-ray (with {\it
Chandra}), and discuss the allowed space density of QSOs in the
early universe.
We adopt a definition of QSOs comprising all objects with
strong, high-ionization emission lines and $M_{145} \leq -21$, 
including both conventional, broad-lined
(type-1) QSOs and narrow-lined, obscured (type-2) QSOs.
Throughout this paper we use 
a cosmology with $h, \Omega_{\rm tot}, \Omega_m, \Omega_{\Lambda} =
0.7, 1.0, 0.3, 0.7$; magnitudes and colors are measured in the AB system.
\section{The Database}
The present work surveys an area of $320$ arcmin$^2$, subdivided in two
fields centered on the {\it Chandra} Deep Field-South (CDF-S) and Hubble Deep
Field-North (HDF-N), each covering $\approx 10'\times 16'$.
The optical data have been obtained with the
Advanced Camera for Surveys (ACS) onboard {\em HST} in the framework of the
GOODS/ACS survey described in \cite{Giava03a}. 
Mosaics have been created from the first three epochs of
observations, out of a total of five, in the bands F435W($B_{435}$),
F606W($V_{606}$), F775W ($i_{775}$), F850LP($z_{850}$).
The catalogs used to select high-$z$ QSOs have been  generated using
the SExtractor software,
performing the detection in the $z_{850}$
band and then using the isophotes defined during this process as
apertures for photometry in the other bands \citep{Giava03a}.

The HDF-N and CDF-S fields have X-ray observations of 2~Ms and
1~Ms, respectively (Alexander et~al. 2003, hereafter A03; Giacconi
et~al. 2002, hereafter G02),
providing the deepest views of the Universe in the 0.5--8.0~keV band.
The X-ray completeness limits over $\approx$~90\%
of the area of the GOODS fields are
similar, with flux limits (S/N$=5$) of $\approx
1.7\times10^{-16}$~erg~cm$^{-2}$~s$^{-1}$ (0.5--2.0~keV) and $\approx
1.2 \times10^{-15}$~erg~cm$^{-2}$~s$^{-1}$ (2--8~keV) in the HDF-N
field, and $\approx 2.2\times10^{-16}$~erg~cm$^{-2}$~s$^{-1}$
(0.5--2.0~keV) and $\approx 1.5\times10^{-15}$~erg~cm$^{-2}$~s$^{-1}$
(2--8~keV) in the CDF-S field (A03).
The sensitivity at the aim point is about 2 and 4 times better for the
CDF-S and HDF-N, respectively.
As an example, assuming an X-ray
spectral slope of $\Gamma=$~2.0, a source detected with a flux of
$1.0\times10^{-16}$~erg~cm$^{-2}$~s$^{-1}$ would have both
observed and rest-frame luminosities of 
$8\times 10^{42}$~erg~s$^{-1}$, and
$3\times 10^{43}$~erg~s$^{-1}$ at $z=3$, and $z=5$,
respectively (assuming no Galactic absorption).  
A03 produced point-source catalogs for the HDF-N and CDF-S and G02 for the
CDF-S. 
%
\section{Selection of the QSO candidates and determination of the redshifts}
The selection of the QSO candidates has been carried out
in the magnitude interval 
$22.45 < z_{850} < 25.25$.
QSO colors in the ACS bands (Fig.~\ref{fig-qsocol}) have been
estimated as a function of redshift using a template \citep{sc90}  
of the QSO spectral energy distribution (SED) convolved with the 
\cite{Madau99} model of the intergalactic medium (IGM) absorption.
Four optical criteria have been tailored in order
to select QSOs at progressively higher redshift in the
interval $3.5 \mincir z \mincir 5.2$:
{\small
\begin{eqnarray}
i-z&<&0.35 ~~{\rm AND}~~ 1.1<B-V<3.0 ~~{\rm AND}~~ V-i<1.0 ~~~~\\
i-z&<&0.35 ~~{\rm AND}~~ B-V>3.0\\
i-z&<&0.5  ~~{\rm AND}~~ B-V>2.0 ~~{\rm AND}~~ V-i>0.8 \\
i-z&<&1.0  ~~{\rm AND}~~ V-i>1.9
\end{eqnarray}
}
They select a broad range of high-$z$ AGN, not limited to broad-lined
(type-1) QSOs, and are less stringent than those typically used to
identify high-$z$ galaxies (e.g., Giavalisco et al. 2003b).
Below $z \simeq 3.5$ the typical QSO colors in the ACS bands move
close to the locus of stars and low-redshift galaxies. 
Beyond $z \simeq 5.2$ the $i-z$ color starts increasing
and infrared bands would be needed to identify QSOs efficiently
with an ``$i$-dropout'' technique.
To avoid contamination from spurious sources we have limited our
selection to $z_{850}$ detections with $S/N > 5$.  The criteria have
been applied independently and produced in total
645 candidates in the CDF-S and 557 in the HDF-N.

Two QSOs were found in the literature within the magnitude and
redshift ranges of interest ($22.45 < z_{850} < 25.25$, $3.5 \leq z
\leq 5.2$): CDF-S $033229.8-275105$($z=3.700$) and HDF-N
$123647.9+620941$($z=5.186$). Both have been selected with the present
criteria
\footnote{
The X-ray emitting
radio galaxy HDF-N $123642.0+621331$, identified by Waddington
et al. (1999) at $z=4.424$, has not been selected.
It should be noted, however, that the redshift of this source
has been questioned by \cite{Barger00}.}.
Ten galaxies with $3.5 \leq z \leq 5.2$ are known in
the HDF-N \citep{Cohen00,Dawson01,Dawson02}. Eight are selected with
the color criteria (1-4) (see Fig.~\ref{fig-qsocol}). The two missed
objects (HDF-N 1236279+6217504 and 1236376+621453) are both
positionally uncertain identifications with a spectrum ``solo-line
with no continuum'' \citep{Dawson01}.  HDF-N 1236279+6217504 has no
optical counterpart in our ACS photometric catalog, and the supposed
counterpart of HDF-N 1236376+621453 has colors very far from our
criteria ($B_{435}-V_{606}=1.4$, {$i_{775}-z_{850}=1.3$,
$V_{606}-i_{775}=2.3$).

The optical candidates selected with the criteria (1-4) have been
matched with X-ray sources detected by {\it Chandra} (A03, G02) within
an error radius corresponding to the $3~\sigma$ X-ray positional
uncertainty. With this tolerance the expected number of false matches
is five and indeed two misidentifications, i.e. cases in which a brighter
optical source lies closer to the X-ray position, have been rejected
(both in the CDF-S). The sample has been reduced in this way to 11
objects in the CDF-S and 6 in the HDF-N (Tab.~\ref{tbl-cand}).
Type-1 QSOs with $M_{145}<-21$, given the measured dispersion in their
optical-to-X-ray flux ratio \citep{Vignali03}, should be 
detectable in our X-ray observation up to $z \magcir 5.2$.
Conversely, any $z>3.5$ source in the GOODS region
detected in the X-rays must harbor an AGN 
($L_x(0.5-2~{\rm keV}) \magcir 10^{43}~{\rm erg~s^{-1}}$).

Photometric redshifts of the 17 QSO candidates 
(Tab.~\ref{tbl-cand}, Col. 11)
have been estimated by comparing with a $\chi^2$ technique (see
\cite{Arnouts99} for details)
the observed ACS colors to those expected on the basis of
{\em a)} a library of template SEDs of galaxies (the ``extended Coleman'' of
\cite{Arnouts99}); 
{\em b)} the QSO SED described in \S~3.
In general the estimates {\em a)} and {\em b)} are similar, 
since the criteria (1-4) both for galaxies and QSOs rely
on a strong flux decrement in the blue part of the spectrum 
- due to the IGM and possibly an intrinsic Lyman limit absorption - 
superimposed on an otherwise blue continuum.
Seven objects out of the 17 selected have spectroscopic
confirmations (Tab.~\ref{tbl-cand}, Col.~12). 
Four are QSOs with redshifts between $2.7$ and $5.2$, in
good agreement with the photometric redshifts based on the QSO SED.
Three are reported to be galaxies and the relatively large offsets 
between the X-ray and optical positions suggest that they could be
misidentifications.
\section{Discussion and Conclusions}
We have compared the QSO counts observed in the $z_{850}$ band 
with two phenomenological and two more physically motivated models
(Fig.~\ref{fig-counts}).
In the absence of a complete spectroscopic follow-up
we focus the comparison 
in the redshift range $4 \leq z \leq 5.2$,
where the selection criteria (1-4) and the photometric redshifts are
expected to be highly complete and reliable.
In addition to one spectroscopically confirmed QSO (HDF-N
$123647.9+620941$, $z=5.186$) we estimate that
between $1-3$ more QSOs at $z > 4$ are present in
the GOODS, depending on whether galaxy or QSO SEDs are adopted for the
photometric redshifts.

{\it Phenomenological models}. The double power-law fit of the 2QZ
QSO LF (Boyle et al. 2000) has been extrapolated for $z>2.7$ (the peak of
QSO activity) in a way to produce a power-law decrease of the number
of bright QSOs by a factor 3.5 per unit redshift interval,
consistent with the $3.0^{+1.3}_{-0.9}$ factor 
found by \cite{fan01b}.
The extrapolation is carried out either as a Pure Luminosity Evolution
(PLE) or a Pure Density Evolution (PDE), assuming that the slopes of the LF
power-laws remain unchanged all the way up to $z \sim 5$.
The PLE model predicts about $17$ QSOs with $z_{850}<25.25$ at
redshift $z>4$ ($27$ at $z>3.5$) in the $320$ arcmin$^2$ of the two GOODS
fields, and is
inconsistent with the observations at a more than $3 \sigma$ level. 
The PDE estimate is $2.9$ QSOs at $z>4$ ($6.7$ at $z>3.5$).

{\it Physically motivated models}.  
It is possible to connect QSOs with DMH formed in hierarchical
cosmologies with a minimal set of assumptions (MIN model, e.g. \cite{haiman01}): 
{\em a)} QSOs are hosted in newly formed halos with
{\em b)} a constant SMBH/DMH mass ratio $\epsilon$ and
{\em c)} accretion at the Eddington rate.
The bolometric LF of QSOs is then:
%
\begin{equation} 
\Phi(L|z)dL =  n_{\rm PS}(M_H|z) \int_{t(z)-t_{\rm duty}}^{t(z)}
\hskip -11mm
P(t_f|M_H,t(z))dt_f \epsilon^{-1} {{dL}\over{L_{\rm Edd}}} 
\label{eq:hh}
\end{equation}
%
where the abundance of DMHs of mass $M_H$ 
is computed using the Press \& Schechter (1974) recipe, 
the distribution $P(t_f|M_H,t(z))dt_f$ of the
formation times $t_f$ follows
\cite{lc93}, $t_{\rm duty}$
is the QSO duty cycle and $L_{Edd}=10^{4.53}~L_{\odot}$ is the Eddington
luminosity of a $1\ M_\odot$ SMBH.
The MIN model is known to overproduce
the number of high-$z$ QSOs \citep{haiman99} and
in the present case predicts $151$ QSOs at $z>4$ ($189$ at $z>3.5$).
Alternatively, QSOs are assumed to shine a $t_{\rm delay}$ time
after DMH formation (DEL model; Monaco, Salucci \& Danese 2000;
Granato et al. 2001).
The QSO LF of eq.~\ref{eq:hh} is then computed at a redshift
$z'$ corresponding to $t(z)-t_{\rm delay}$.  The predictions of the
DEL model are very close to the PDE, with $3.2$ QSOs expected at
$z>4$ ($4.8$ at $z>3.5$), in agreement with the observations.
While all the models presented here are consistent with the recent QSO LF 
measurement of the COMBO-17 survey \citep{COMBO17} at $z \simeq 2-3$, 
only the PDE and DEL models fit the COMBO-17 LF in the range
$4.2<z<4.8$ and $M_{145}<-26$.

At $z>4$ the space density of moderate luminosity ($M_{145} \simeq
-23$) QSOs is significantly lower than the prediction
of simple recipes matched to the SDSS data, such as a PLE evolution
of the LF or a constant universal efficiency in the formation of SMBH
in DMH.
A similar result has been obtained at $5 \mincir z \mincir 6.5$, by
\cite{Barger03} who also showed that the QSO contribution to the UV
background is insufficient to ionize the IGM at these
redshifts.
A flattening of the observed high-$z$ LF 
is required below the typical luminosity regime 
($M_{145} \mincir -26.5$) probed by the
SDSS.
This is an
indication that at these early epochs the formation or the feeding of
SMBH is strongly suppressed in relatively low-mass DMH, as a
consequence of feedback from star formation \citep{granato01} and/or
photoionization heating of the gas by the UV background
\citep{haiman99}.
\acknowledgments
We are grateful to H.Spinrad, S.Dawson, and A.Dey for providing
the redshift for HDF-N $123627.5+621158$ and to C.Steidel for the
spectrum of the galaxy HDF-N $123644.1+621311$.
The work of DS was carried out at the Jet Propulsion Laboratory,
California Institute of Technology, under a contract with NASA.
DMA, FEB, and WNB thank the NSF CAREER award AST-9983783. DMA
also acknowledges the support provided by a Royal Society
University Research Fellowship.
We thank STScI grant HST-GO-09425.26-A (FEB, WNB),
HST-GO-09425.13-A (CMU, EC, ET), and ASI 
grant I/R/088/02 (SC,
AG, MN, EV).

%
\begin{deluxetable}{cccccccccccc}
\tabletypesize{\tiny}
\tablecaption{High-redshift QSO candidates in the CDF-S. \label{tbl-cand}}
\tablewidth{0pt}
\tablehead{
\multicolumn{2}{c}{Optical} & \multicolumn{2}{c}{Optical -- X-ray} &
\colhead{$z_{850}$} &
\colhead{$B-V$} & \colhead{$i-z$} 
& \colhead{$V-i$} &\colhead{$F_x$\tablenotemark{\P}} & \colhead{$F_x$\tablenotemark{\P}}
& \colhead{phot.} & \colhead{spectr.} \\
\colhead{RA} & \colhead{DEC} & \colhead{$\Delta$ RA} &
\colhead{$\Delta$ DEC} &
\colhead{AB} & \colhead{AB} & \colhead{AB} & \colhead{AB} &
\colhead{$0.5-2~{\rm keV}$} &\colhead{$2-8~{\rm keV}$}&
\colhead{redshift} & \colhead{redshift} \\
\colhead{$3^h+ ~m~ s$} & \colhead{$-27^{\circ}+ ~'~``$} & 
\colhead{(arcsec)} & \colhead{(arcsec)} & 
(mag) & (mag) & (mag) & (mag) & \multicolumn{2}{c}{$(10^{-16} {\rm
erg~s{-1}~cm^{-2}})$} &  
\colhead{gal, QSO \tablenotemark{\S}}& \colhead{ } 
}
\startdata
 32 03.03& 44 50.1& $+0.0$ & $+0.5$ & 24.77& $1.39$& $0.35$ & 0.74 & $6.84$&$8.99$& 2.5,3.3 & \\
 32 04.94\tablenotemark{\dagger}& 
           44 31.7& $+1.7$ & $-0.5$ & 23.65& $2.01$& $0.03$ & 0.20 & $<0.74$ &$4.10$& 3.7,3.7 & 3.462\tablenotemark{a}\\
 32 14.44& 44 56.6& $-0.5$ & $-0.3$ & 23.05& $2.12$& $0.44$ & 1.46 & $0.41$&$<4.3$  & 0.7,4.3 &\\
 32 18.82& 51 35.3& $-0.9$ & $+0.1$ & 24.82& $2.52$& $0.35$ & 0.49 & $0.74$&$14.0$& 3.9,3.8 &\\
 32 19.40\tablenotemark{\dagger}& 
           47 28.3& $+0.8$ & $-0.7$ & 24.60& $1.30$& $0.07$ & 0.45 & $<0.31$ &$1.17$& 3.4,3.5 &\\
 32 29.29\tablenotemark{\ddag}& 
           56 19.3& $-1.0$ & $+0.1$ & 25.05& $>3$  & $0.12$ & 1.65 & $0.50$&$<7.6$& 4.6,4.7 &\\
 32 29.85& 51 05.8& $-0.1$ & $+0.1$ & 24.59& $2.25$& $0.01$ & 0.59 & $3.06$&$31.8$& 3.7,3.8 & 3.700\tablenotemark{b}\\
 32 39.67& 48 50.6& $-0.1$ & $+0.1$ & 24.55& $2.96$& $0.24$ & 0.97 & $7.48$&$70.6$& 3.9,4.0 &\\
 32 41.87& 52 02.5& $-0.3$ & $+0.1$ & 22.47& $1.24$& $0.22$ & 0.46 & $16.7$&$38.2$& 0.5,3.6 &\\
 32 42.84& 47 02.5& $+0.0$ & $-0.1$ & 24.91& $1.28$& $0.10$ & -0.04& $6.31$&$16.4$& 3.5,3.4 &\\
 32 50.25& 52 51.9& $+0.3$ & $-0.2$ & 25.19& $1.54$& $0.10$ & 0.32 & $24.0$&$75.3$& 0.6,3.6 &\\
\tableline
\\
\multicolumn{12}{c}{\small High-redshift QSO candidates in the HDF-N.}\\
 {$12^h+ ~m~ s$} & {$62^{\circ}+ ~'~``$} \\ 
\tableline
\\
 36 27.59& 11 58.7& $-1.1$ & $+0.3$ &23.60& $1.28$& $0.20$ & 0.69 & $1.06$&$22.0$& 0.4,3.7 &0.395\tablenotemark{c}\\
 36 42.21& 17 11.6& $-0.1$ & $+0.1$ &24.01& $1.12$& $0.22$ & 0.59 & $9.48$&$23.2$& 2.8,2.9 &2.724\tablenotemark{d}\\    
 36 43.09& 11 08.9& $-0.1$ & $+0.9$ &23.13& $1.15$&$-0.04~$& 0.54 & $0.75$&$<2.0$& 0.4,3.5 &0.299\tablenotemark{e}\\
 36 44.10\tablenotemark{\dagger}& 
           13 11.0& $+0.1$ & $-1.1$ &23.87& $1.35$& $0.06$ & 0.45 & $<0.26$ & $<1.4$ & 3.4,3.6 &2.929\tablenotemark{f}\\
 36 46.07& 14 49.2& $+0.1$ & $+0.3$ &25.16& $1.36$& $0.25$ & 0.99 & $<0.34$ & $<1.4$ & 0.6,3.8 &\\
 36 47.96& 09 41.6& $+0.1$ & $+0.3$ &23.72& $4.45$& $0.13$ & 2.12 & $2.72$&$4.89$& 4.7,4.8 &5.186\tablenotemark{g}
\enddata
\tablenotetext{\P}{\scriptsize Objects with upper limits
both in the soft and hard band have been detected in other sub-bands.}
\tablenotetext{\S}{\scriptsize The first photometric redshift is computed on the
basis of galaxy SEDs, the second for QSO SEDs.}
\tablenotetext{\dagger}{\scriptsize From the supplementary X-ray
catalog by A03.} 
\tablenotetext{\ddag}{\scriptsize Detected in the X-ray by G02 and not by A03. 
X-ray coordinates and fluxes derived from G02. Positions in G02 are typically offset $+0.07^s$
in RA and $-1''$ in DEC with respect to the optical ACS positions. The
correction has been applied in Cols.3-4.}
\tablenotetext{a}{\scriptsize QSO showing relatively narrow emission
lines with a P-Cyg profile \citep{sc00}.}
\tablenotetext{b}{\scriptsize Type-2 QSO \citep{Norman02}.}
\tablenotetext{c}{\scriptsize Galaxy: H.~Spinrad, private communication.}
\tablenotetext{d}{\scriptsize Broad emission-line QSO \citep{Barger01}.}
\tablenotetext{e}{\scriptsize Galaxy identified by \cite{Cohen00}.} \tablenotetext{f}{\scriptsize Galaxy identified by \cite{dickinson98}. 
No spectroscopic evidence for AGN activity 
is present in the spectrum.} \tablenotetext{g}{\scriptsize QSO identified by \cite{Barger02}.}
\end{deluxetable}
\begin{figure}
\epsscale{.60}
\plotone{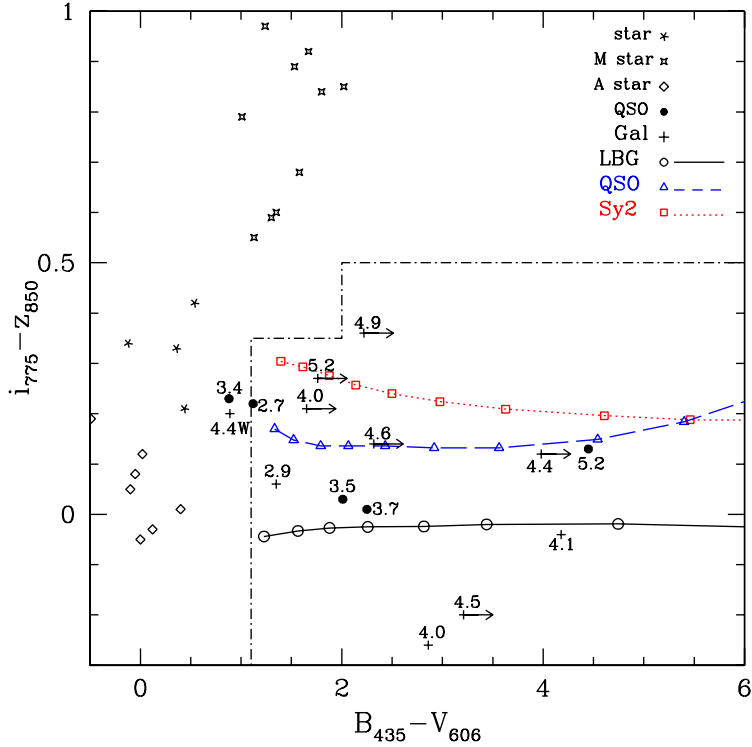}
\caption{Expected $i_{775}-z_{850}$ vs. $B_{435}-V_{606}$ colors of
QSOs as a function of redshift compared to other classes of objects. 
The dashed line shows the locus of QSOs at $z\geq 3.5$
estimated with the composite spectrum of CV90.
The continuous and dotted lines show the colors  
of Ly-break and Sy 2 galaxies, derived from the
SEDs of \cite{Schmitt97} and \cite{Arnouts99}, respectively.
The corresponding symbols - open triangles, circles and squares
- start on the left at $z=3.5$
and move to the right in steps of $\Delta z=0.1$. The observed colors
of five QSOs are marked with filled circles and the corresponding
redshifts. The same is done for ten high-$z$ galaxies identified with
crosses. The radio galaxy HDF-N $123642.0+621331$ is
marked with the label $4.4$W.
The positions of various types of stars is also shown.
The dot-dashed line represents the projection of the selection criteria 1) + 3).
\label{fig-qsocol}}
\end{figure}
\begin{figure}
\epsscale{1.0}
\plottwo{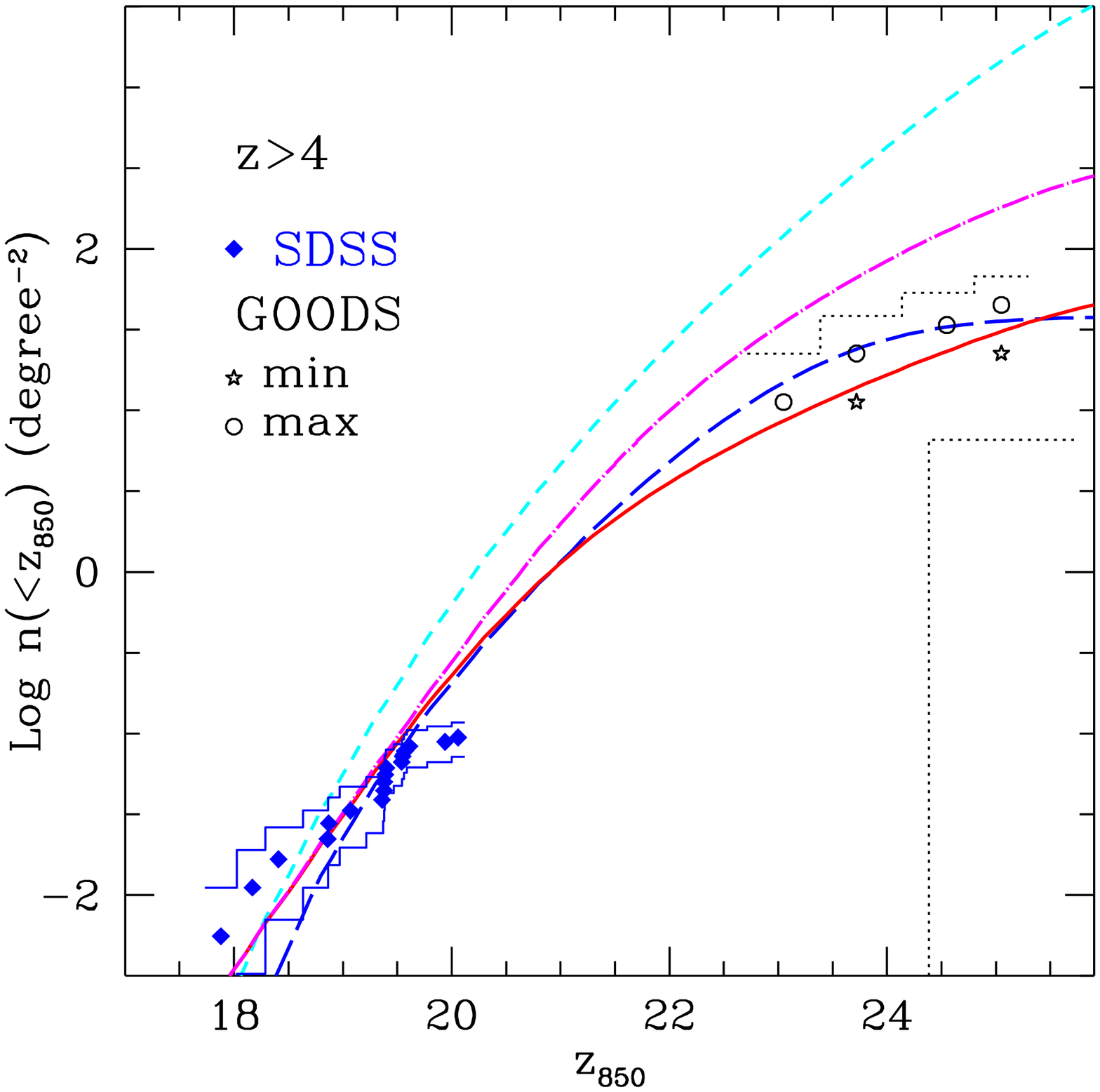}{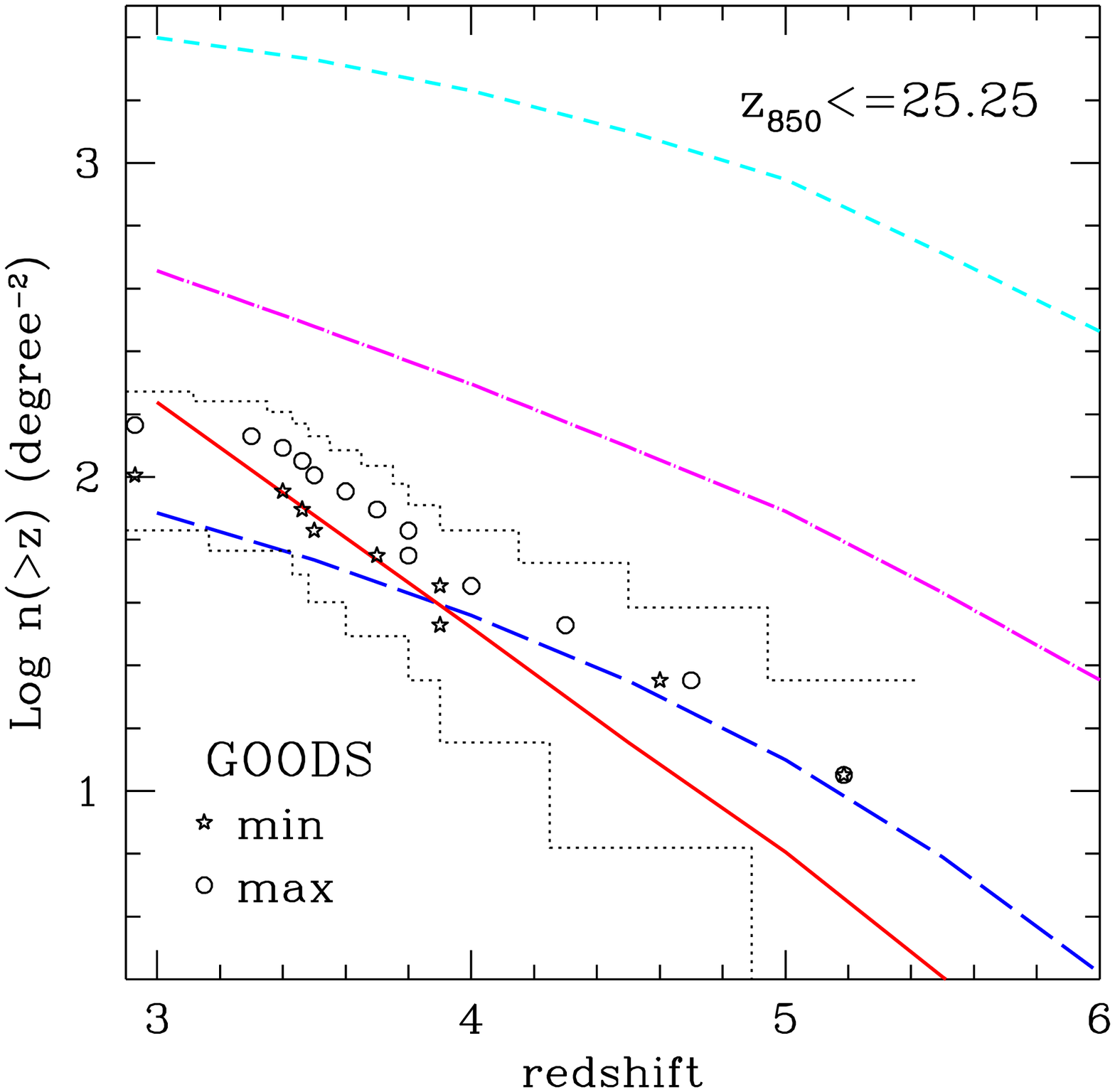}
\caption{Comparison of the observed QSO counts and redshift
distribution with model predictions.
Circles and stars show the ``maximal'' and ``minimal'' estimates of
the GOODS counts, respectively (see text). 
The dotted segments show the corresponding $1~\sigma$ upper (maximal case)
and lower (minimal case) confidence limits.
Models are represented by four smooth lines: dot-dashed PLE, 
continuous PDE, short-dash MIN \citep{haiman01}, 
long-dash delayed QSO shining.
A value of $\sigma_8=0.83$ has been assumed in the MIN and DEL models,
whose parameters have been fixed in order to reproduce the
QSO LF at $z\sim3$. 
In the MIN model: $\log \epsilon=-3.2$ and $t_{\rm duty}=0.015\ {\rm Gyr}$. 
In the DEL model: $\log \epsilon=-3$ and $t_{\rm duty}=0.013\
{\rm Gyr}$; the delay time $t_{\rm delay}(M_H)$ is set to $0.75$ Gyr above
$10^{12.5}\ M_\odot$ and is decreased as $\propto M_H^{-0.7}$ at lower
masses \citep{granato01}. As dynamical and cooling times are shorter at high
redshift, $t_{\rm delay}$ is assumed to decrease as
$(1+z)^{-2}$ above $z=2.7$.
\label{fig-counts}
}
\end{figure}
\end{document}